# Illuminating Invisible Grain Boundaries in Coalesced Single-Orientation WS$_2$ Monolayer Films

*Danielle Reifsnyder Hickey,[1] Nadire Nayir,[2,3,4] Mikhail Chubarov,[2] Tanushree H. Choudhury,[2] Saiphaneendra Bachu,[1] Leixin Miao,[1] Yuanxi Wang,[2] Chenhao Qian,[1] Vincent H. Crespi,[2,5] Joan M. Redwing,[1,2] Adri C.T. van Duin,[2,3] Nasim Alem[1]*

[1]Department of Materials Science and Engineering, The Pennsylvania State University, University Park, PA 16802, USA

[2]2D Crystal Consortium-Materials Innovation Platform (2DCC-MIP), Materials Research Institute, The Pennsylvania State University, University Park, PA 16802, USA

[3]Department of Mechanical Engineering, The Pennsylvania State University, University Park, PA 16802, USA

[4] Department of Physics, Karamanoglu Mehmet University, Karaman, 70000, Turkey

[5]Department of Physics, The Pennsylvania State University, University Park, PA 16802, USA

**Abstract**: Engineering atomic-scale defects is crucial for realizing wafer-scale, single-crystalline transition metal dichalcogenide monolayers for electronic devices. However, connecting atomic-scale defects to larger morphologies poses a significant challenge. Using electron microscopy and atomistic simulations, we provide insights into WS$_2$ crystal growth mechanisms, providing a direct link between synthetic conditions and the microstructure. Dark-field TEM imaging of coalesced monolayer WS$_2$ films illuminates defect arrays that atomic-resolution STEM imaging identifies as translational grain boundaries. Imaging reveals the films to have nearly a single orientation with imperfectly stitched domains. Through atomic-resolution imaging and ReaxFF reactive force field-based molecular dynamics simulations, we observe two types of translational mismatch and discuss their atomic structures and origin. Our results indicate that the mismatch results from relatively fast growth rates. Through statistical analysis of >1300 facets, we demonstrate that the macrostructural features are constructed from nanometer-scale building blocks, describing the system across sub-Ångstrom to multi-micrometer length scales.

**Main Text**: Transition metal dichalcogenides (TMDs) represent a frontier of semiconductor research due to their exciting optoelectronic properties and potential for integration into very thin devices.[1–9] However, incomplete understanding of how to achieve single-crystalline, device-scale films with minimal atomic-scale defects[10–12] has limited their performance.[13–16] To date, the materials properties of TMDs have been explored largely using *isolated* flakes that have either been exfoliated from bulk materials[17–19] or grown by chemical and physical deposition techniques.[20–22] Recently, growth of coalesced monolayer films on the wafer scale[23–30] has been achieved by more controlled growth methods, such as metal–organic chemical vapor deposition (MOCVD), in which gas-phase precursors provide increased tunability, and CVD enhanced by using multiple temperature zones. As achieving large-area monolayer films becomes possible, it is more critical to uncover defects, such as grain boundaries or point defects that arise from synthesis, and to understand the underlying physics behind their formation. With this understanding, we can further engineer such defects and alleviate obstacles to single crystallinity.[31–33]

Typically, islands of TMDs are reported to have triangular or hexagonal morphologies following the kinetic Wulff construction.[10,34–37] In such cases, the steady-state flake morphology can be described as a function of the thermodynamic conditions during synthesis, such as the chemical potentials of the elements involved, determined by precursor vapor pressures.[38] Therefore, continuous epitaxial films resulting from the coalescence of orientationally aligned islands are also expected to exhibit facets along high-symmetry directions. Yet, the transmission electron microscopy (TEM) investigation presented here for 2H-$WS_2$ films grown by MOCVD on 2" c-plane sapphire [(0001) α-$Al_2O_3$] substrates shows strikingly different grain boundary structures compared to grain boundaries arising from hexagonal symmetry.

Initial bright-field TEM imaging and selected-area electron diffraction (SAED) of the as-grown $WS_2$ films that have been transferred from the growth substrate onto TEM grids show an almost uniform and featureless nearly single-crystalline film across multi-micron regions (Fig. 1a). This observation directly corresponds with the results of measurements such as AFM and in-plane X-ray diffraction that yield information across larger length scales, which also support the monolayer thickness and large-area, millimeter-scale epitaxial nature of the $WS_2$ grown on sapphire.[29] However, new features are uncovered when using dark-field (DF-) TEM imaging. We observe unexpected, irregular, linear defects throughout the several-micron areas of the almost epitaxial monolayer film that possesses a nearly single-crystalline diffraction pattern (Fig. 1b). DF-TEM imaging further uncovers that the monolayer consists of two types of sub-µm regions, shown in two different shades of gray. This variation in the diffraction contrast suggests a subtle offset between the neighboring grains, while the grains still maintain a single-crystalline diffraction pattern.

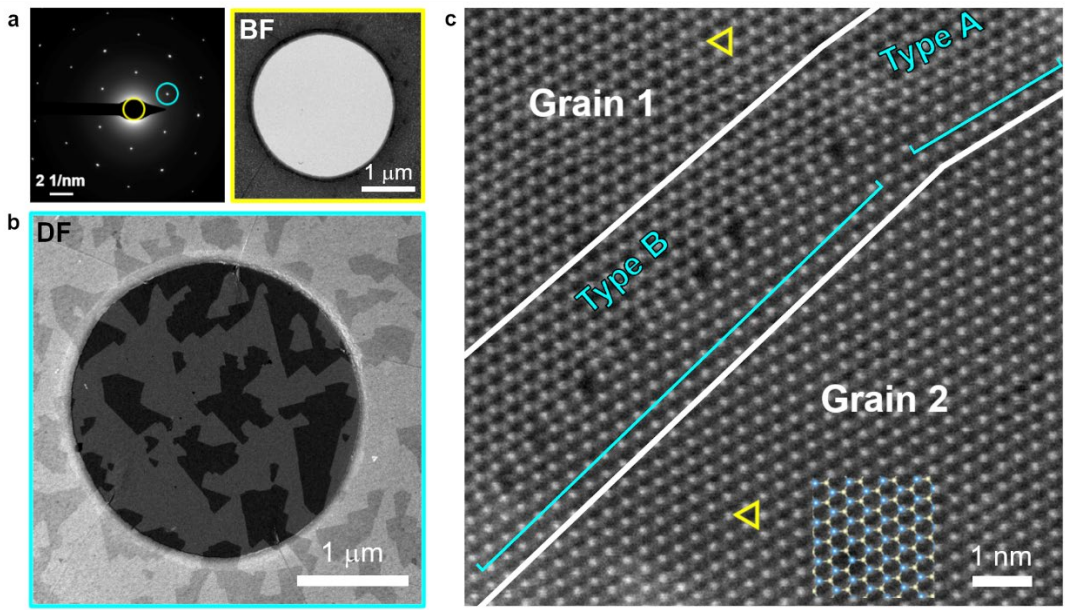

Fig. 1. Structure of nearly single-orientation monolayer 2H-WS$_2$ film and its GB defects. (a) SAED pattern (left) showing an apparently single-crystalline morphology, which corresponds to the featureless monolayer area shown in the BF-TEM image (right, created from the direct beam indicated by the yellow circle in the SAED pattern). The circle in the BF-TEM image represents the free-standing film area in a hole in the Quantifoil carbon TEM substrate. (b) DF-TEM image (created from the Bragg spot indicated by the cyan circle in Fig. 1a) of the area in the BF-TEM image in (a), showing irregular, faceted features in the monolayer. (c) Atomic-resolution HAADF-STEM image of two similarly oriented grains meeting at a GB (in the atomic model, blue represents W, and yellow represents S).

Further inspection of atomic-resolution HAADF-STEM imaging (Fig. 1c) reveals that two neighboring WS$_2$ grains (corresponding to two different shades of gray in the DF-TEM image) both retain the 2H structure and the same in-plane crystallographic orientation across the grain boundary (GB). Despite the SAED pattern appearing nearly single-crystalline, the atomic-resolution imaging of the GB separating the regions with the two shades of gray in DF-TEM shows that these WS$_2$ films contain distinct GB atomic structures. Strikingly, no in-plane rotation of the grains is observed at this GB (confirmed by atomic resolution imaging in Fig. 1c and SAED in Fig. 1a), and the WS$_2$ lattice orientation is nearly identical on both sides of the GB (marked by the sulfur column triangles on Fig. 1c). It is important to note that the GB structures reported here are distinct from mirror twin boundaries (GBs existing between two 2H grains with 60° relative rotation) that have been reported to show two similar types of contrast in DF-TEM for monolayer TMDs.[39] Instead, we observe irregular, faceted angles that do not adopt the characteristic TMD island shapes with hexagonal symmetry. High-resolution HAADF-STEM imaging of the GBs indicates a sub-unit cell translational offset between the neighboring epitaxial grains with no apparent in-plane grain misorientation. In the nearly single-orientation films, the observed translational GBs create a variety of defect arrays arising from sub-unit-cell translational disregistry (either transverse or longitudinal) between epitaxially oriented grains and can lead to unexpected structural ordering throughout the film.

The widespread GB defect structures shown in Fig. 1 (see also Fig. S1) represent regions where nuclei have coalesced with two different types of mismatch along these boundaries (Fig. 2). In the first case (type A), a translational grain boundary is created when the hexagonal lattice is compressed perpendicular to the GB, resulting in decreased metal-to-metal distances while retaining the original hexagonal symmetry (3|W) of the grain interiors (Fig. 2a-c). In this type of boundary, although the symmetry is approximately maintained at the GB, the boundary atoms are significantly compressed towards each other, creating a compressive strain (Fig. 2c). In the second case (type B), a translational GB is created when the hexagonal lattices meeting at the GB undergo a sub-unit-cell shift *parallel* to the GB, leading to a transverse disregistry (Fig. 2a,b,d). This transverse lattice disregistry leads to broken hexagonal symmetry, forming through shear a rectangular arrangement of tungsten atoms (4|W) connected by sulfur atoms along the GB.

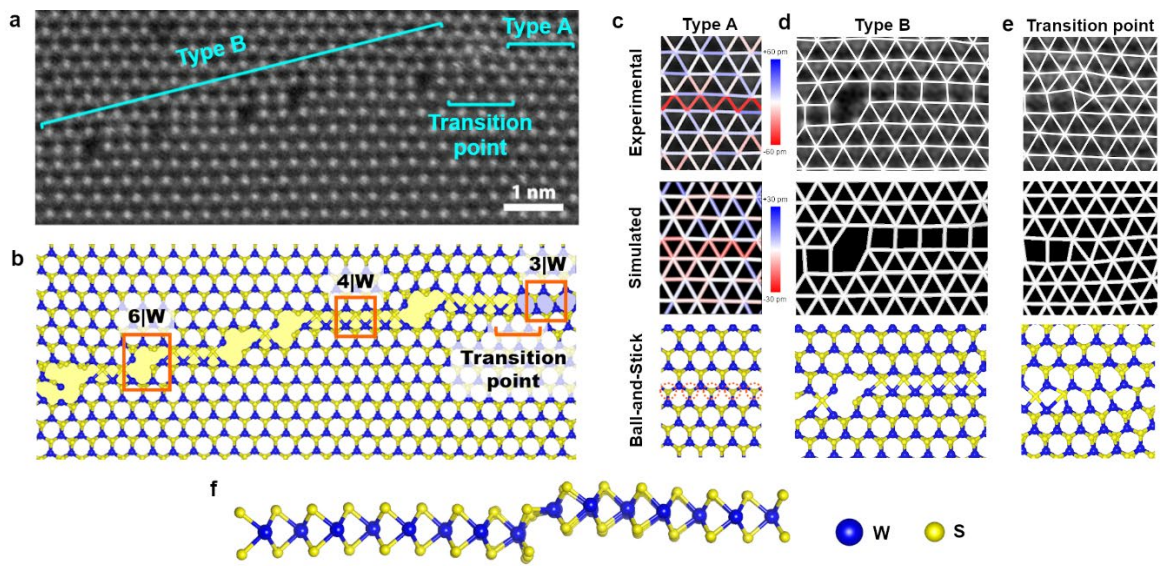

Fig. 2. Local structures obtained from experimental HAADF-STEM imaging and ReaxFF MD simulation. (a) HAADF-STEM image showing type A and B GBs, as well as the transition point between them. (b) ReaxFF MD simulation equilibrated at 300K of the structure shown in (a). (c-e) Comparison (experiment, simulation, and ball-and-stick atomic model) of the grain boundary structures between adjacent epitaxial, oriented grains of $WS_2$: (c) type A mismatch, in which a single vacancy line results in compression of the planes (red) but retains the hexagonal arrangement (3|W) of metal atoms; (d) type B mismatch, in which a sub-unit-cell offset between the grains results in a rectangular arrangement (4|W) of metal atoms at the GB; and (e) the transition point between GB regions with 3|W and 4|W. Each dashed orange circle on the ball-and-stick model in (c) represents an S vacancy. (f) Cross-sectional view of the ReaxFF MD simulated model showing the abrupt atomic displacement at the type B GB.

ReaxFF reactive force field molecular dynamics (MD) simulations, as depicted in Fig. 2b-f, support the formation of the type A and B GB structures. ReaxFF enables large-scale (>1000 atoms) simulations on material chemistry and has been developed for a wide range of materials, including 2D materials.[40,41] Specifically, the type A structure can form with the presence of sulfur vacancies along the GB. In this case, sulfur vacancies (removal of one S atom at an S sublattice site) assemble into a single vacancy line defect (Fig. 2c). The three under-coordinated W atoms surrounding every S vacancy then relax toward the single-S-atom columns and result in the compression of W–W distance projections across the type A GBs. Additionally, simulation shows that the formation of the single vacancy line drives a slight out-of-plane buckling of the GB by splitting the double-stacked S atoms. This buckling can create a mistilt along the z-direction and result in differences in the HAADF-STEM intensities of the S atoms in the two grains (Fig. 1c), which is a well-known effect of sample tilt relative to the incident electron beam.[42,43] Although the type A boundary observed here results from the film growth conditions (see Fig. S2), structurally it resembles similar compression lines that have previously been demonstrated to form in TMDs under high-energy electron beams.[44–46] However, here they originate from the sub-unit-cell offset present during coalescence, leading to reduced W-W bond distances at the GB, and further leading to compressive strain. In contrast to type A, the type B GB (Fig. 2d,f) does not arise from the presence of S vacancies but instead represents the atomic configuration where two growing $WS_2$ grains contain sub-unit-cell mismatch parallel to the GB, making it challenging for the grains to stitch together seamlessly. As a result of this shift, the hexagonal symmetry is broken, and the type B GB adopts a rectangular arrangement of W atoms (4|W) along the boundary to account for the mismatch. Type B segments also contain 6|W rings that create steps along the length of the GB. In type B regions, the mismatched lattices adopt a structure more akin to an extended one-dimensional defect or "translational grain boundary," related to those previously reported in graphene.[47–49] Fig. 2f illustrates the out-of-plane distortion that ReaxFF simulation predicts at the type B GB in the case of films that have been released from their sapphire substrates (as investigated here). At these boundaries, two grains with inequivalent zigzag edges (S-terminated and W-terminated) meet with geometry influenced both by the lattice offset and their dissimilar van der Waals interactions with the sapphire substrate during growth.

We experimentally observe transitions between type A and B translational grain boundaries indicating regions with *both* longitudinal *and* transverse shifts (Fig. 2e). According to the MD simulations, transition between types A and B occurs when the 3|W ring (type A) evolves into the 4|W rectangular geometry (type B), associated with a relative shift between the grains parallel to the zigzag edge, by nearly half of the W–W bond distance ($\Delta L = d_{w-w}/2 \approx 1.5$ Å), as well as a lattice distortion of ~0.1 Å perpendicular to the zigzag edge. This shift further eliminates the longitudinal translational offset across the 4|W regions of the type B GB along the zigzag edge, and fully transforms the type B GB into the type A with only transverse translational registry between the two grains. Importantly, the 2H lattice orientation is identical and remains the same for the grains across such boundaries in our experiments. Previously, TMD monolayers have been reported to produce contrast differences in DF-TEM images when they either contain 60°-rotated grains connected by mirror twin boundaries[39] or a phase transition between the 2H and 1T TMD polymorphs.[50] We emphasize here that neither is the case for our system (see Figs. 1c, S3, and S4). In the $WS_2$ monolayers studied here, the rotational mismatch between the two grains is negligible (Fig. 1c), and therefore no mirror twin boundary is observed. We also observe distinct S columns visible on both sides of the GB, ruling out phase transition to 1T (Figs. 1c and 2a). Here, the grains form well-defined GB structures that compensate for slight in-plane offsets of the identically oriented grains, instead of forming a mirror twin boundary with 60° misorientation. While the formation of mirror twin boundaries has been a limiting factor for film growth, the films studied here possess a single orientation across numerous multi-micron areas sampled, despite the observed subtle defect array structures that represent coalescence of sub-micron-scale grains. This achievement underscores a remarkable step forward in large-scale TMD monolayer synthesis.[29]

Despite the long-range geometrical irregularity of the two monolayer regions identified by DF-TEM imaging (Fig. 1b), the two GB structures described in Fig. 2 provide a consistent explanation for the faceted DF-TEM features observed across multi-micrometer areas. Fig. 3 shows the correlation of DF-TEM facets with the atomic GB structures observed *via* HAADF-STEM imaging (see also Fig. S5). Here, the atomic building block registries lead to the type A translational GBs (Fig. 3a, red), which follow the zigzag edge of the 2H crystal structure, and the type B "slanted" translational GBs (Fig. 3a, blue), which deviate from the zigzag edge as a result of the stabilizing 6|W polygonal rings at various periodicities. The type B GB is experimentally observed to occur (Fig. 3a,b) with 6|W units occurring as infrequently as every sixth polygon (creating calculated angle $\theta_{6,1}=7.59°$ with the zigzag edge) and as frequently as every second polygon ($\theta_{2,1}=19.11°$), which creates the maximum possible angle away from the zigzag edge using the 4|W and 6|W components. Experimentally, the average slanted angle is ~15°, which indicates that the type B boundary primarily exists with combinations of the commonly observed $\theta_{2,1}=19.11°$, $\theta_{3,1}=13.90°$, and $\theta_{4,1}=10.89°$. If we consider an average ± 15° off the zigzag edge, a number of possible facet geometries become possible, several of which are illustrated in Fig. 3c. Here, the 60° angles in the red triangle (an ideal flake) represent a standard TMD flake with the three zigzag edges that enclose an equilateral triangle. Blue lines represent facets at ~15° angles to their adjacent red zigzag edges. Further, combinations of these facets illustrate several angles, such as 75°, 165°, and 90°, that are experimentally observed in the grain boundaries studied here.

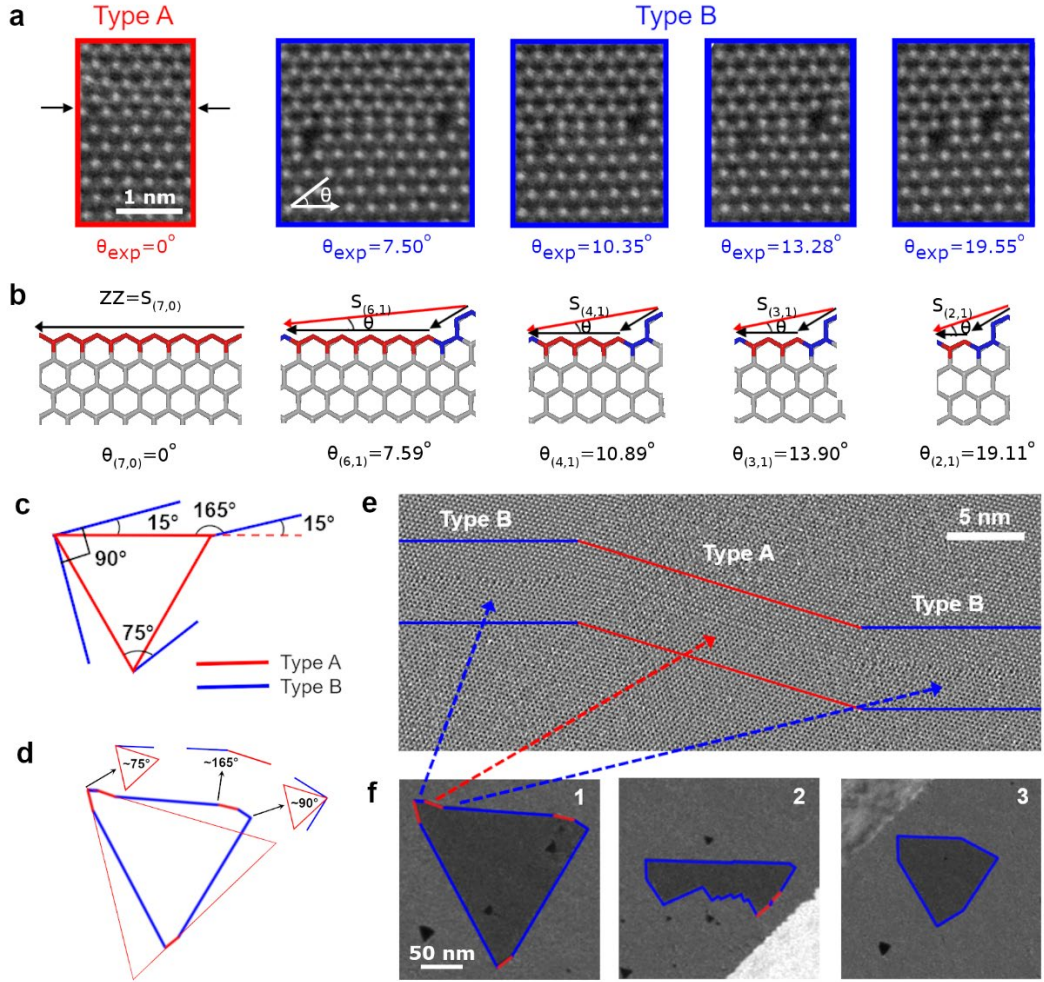

Fig. 3. Correlation of atomic GB structures with facets observed *via* DF-TEM. (a) Experimental HAADF-STEM images of the building block type A and B GB structures with the corresponding experimentally measured angles off the zigzag edge. (b) ReaxFF models corresponding to the images in (a), marked with the angle off the zigzag edge that a periodic stretch of this structure would create. (c) A schematic showing the regular hexagonal symmetry of the type A structure along the zigzag edge (red) and various angles that form from the average type B angle off the zigzag edge (~15°) in combination with type A and type B orientations. (d) An overlay of the hexagonal symmetry of the zigzag edge (red triangle) and an experimentally observed configuration (blue and red polygon, in which the red regions coincide with the predicted zigzag edge and the blue regions are <20° off the zigzag edge. (e) A bandpass-filtered HAADF-STEM image showing connections between three different facets that have type B–type A–type B atomic structures. The facets in this region consist of two boundary morphologies shown as red and blue facets in the DF-TEM image below. (f) Three examples of DF features with the GB facets marked in red or blue according to whether they are type A or type B.

Fig. 3d applies the illustration in Fig. 3c to an experimentally observed DF-TEM image (shown in Figure 3f-1). Here, the red equilateral triangle represents the ideal type A boundary along the zigzag edge, and blue represents the slanted type B boundary. The experimentally obtained red and blue faceted polygon is positioned such that the red facets are overlaid onto the equilateral triangle model. The experimental red facets clearly follow the modeled zigzag edges, and the blue facets all create angles that can be represented by ~10°-20° angles with the red facets. Fig. 3e shows bandpass-filtered HAADF-STEM imaging of two ~165° angles made between type B (blue) and type A (red) facets. These facets correspond to the three indicated edges of DF feature 1 in Fig. 3f-1, in which both red and blue facets are observed, but blue facets are clearly more prevalent. Across numerous GBs analyzed *via* DF-TEM and HAADF-STEM imaging (see also Figs. S6 and S7), including those in DF features 1-3 presented here in Fig. 3f, the type B (blue) structure dominates. Subsequently, the prevalence of blue type B facets clearly provides an explanation for the origin of the irregular GB angles that deviate from the regular 60°/120° symmetry expected for TMD flakes enclosed by zigzag edges.

In fact, large-area, composite DF-TEM mapping (Fig. 4a) including >1300 facets shows the dominance of the irregularly oriented GBs studied here. These boundaries possess an angular distribution (relative to the zigzag edge as identified by HAADF-STEM imaging; see Fig. S7a) that peaks at type B orientations, i.e., angles that are ~15° off the zigzag edge (Fig. 4b). This statistical analysis is consistent with local observations (e.g., Fig. 3e) that the type B GB at ~15° degrees off the zigzag direction is dominant. Nevertheless, a minority population does exist at 0° and 60°, which represents the observed type A GBs. The ReaxFF MD calculations of edge formation energies, computed as a function of the excess sulfur chemical potential, $\Delta\mu_S$, by Eqs. S1, S2, and S4, predict that the slanted edges lose their stability with increasing angle off the zigzag edge and associated kink (6|W) concentration (Figs. 4c and Eq. S5). Additionally, as described in the literature,[51] there is a nearly linear relationship between the slanted angle and growth rate (Fig. 4d and Eqs. S5-S7) that is determined by the formation energy (Eq. S6) necessary to add atoms onto a given edge (Fig. 4e,f). In an S-rich environment such as that experimentally studied here, the high energy cost of 6.40 eV required for the addition of the first zigzag unit (3W+4S atoms) onto the zigzag edge, ZZ, creates an obstacle to its growth (Fig. 4e) because it is considerably larger than the thermal energy, kT ~ 0.1 eV at 1000 °C, as defined by the equation $R(\theta) \approx c_{ZZ}(\theta)e^{-6.40/kT}$ and reported in the earlier work.[51] Therefore, although ZZ is the most stable edge structure, its growth rate is expected to be slowest. In contrast, $S_{(2,1)}$, with a calculated angle of 19.1°, is determined to be a less stable edge structure but to have the highest growth rate. This results from its negligible formation energy of 0.17 eV, which is required for the attachment of an extra zigzag unit (2W+4S atoms) to the 6|W kink of the $S_{(2,1)}$ edge (Fig. 4f), because it has the largest kink density considered here. The details of the theoretical calculations are presented in the SI.

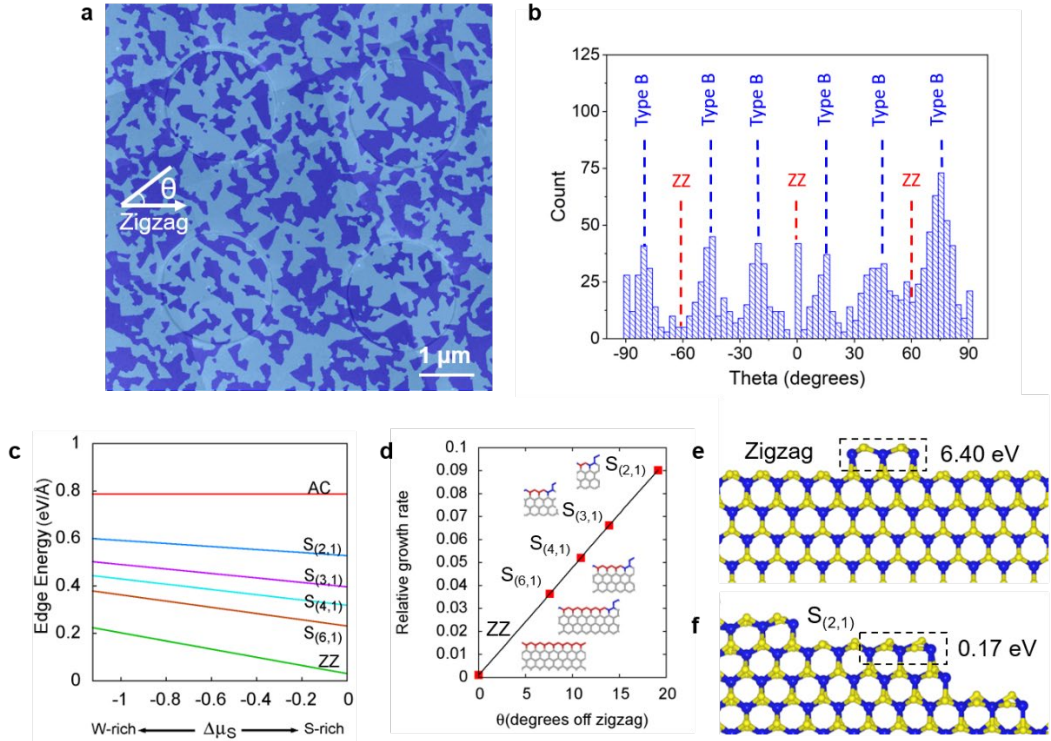

Fig. 4. Distribution of GBs and energetics of GB formation. (a) Large-area false-colored composite DF-TEM map. (b) Histogram of facet orientations measured from the composite DF image in (a) (to reduce error, angles were only measured from free-standing film regions). Zigzag orientations are marked (present but a minor contribution), and type B orientations ~ +/- 15° off the zigzag dominate. (c) ReaxFF-calculated edge formation energies of zigzag (ZZ), slanted (S), and armchair (AC) edges as a function of the excess sulfur chemical potential, $\Delta\mu_S$ (in eV). (d) Calculated relative growth rate of the slanted edge structures with respect to the ZZ edge as a function of slanted angle, $\theta$. $S_{(2,1)}$, $S_{(3,1)}$, $S_{(4,1)}$, and $S_{(6,1)}$ are the slanted edges characterized by the translational vector (n,1) of the 2H-WS$_2$ lattice, where n is the number of the zigzag units along the $a_1$ direction, and the $a_2$-component vector contains only one zigzag unit in the GB of interest, as shown in Fig. S5. (e,f) Illustration of the formation energy for adding atoms (enclosed in the dashed rectangular boxes) onto the reference ZZ and $S_{(2,1)}$ slanted edges, respectively.

This analysis indicates that the growth of the ZZ edge requires near-equilibrium growth conditions and a long duration to complete its linear formation. This is consistent with simulations for graphene[52] and carbon nanotubes,[53] which have established that their zigzag edges are more stable than edges 0°–20° off the zigzag that have growth rates related to the density of kinks. Therefore, for WS$_2$, although the slanted edges are less stable edge structures and should disappear quickly during the growth process, kinetically controlled growth conditions may not allow flakes to complete their zigzag linear formation. Thus, grains with slanted edges may survive during the growth to meet at GBs, as observed in these films (e.g., Fig. 1c), aided by the limited distance between adjacent islands of $WS_2$ during growth. In fact, precedent exists in *isolated* flakes of graphene[51,54] and GaSe[55] for such slanted edges: in a series of growth–etching–regrowth studies, off-zigzag edges (up to 19°) were preferentially observed as faster growing/etching edges than the zigzag edges. These related examples, however, did not have the spatial constraint of nearby islands during edge formation.

Fig. 4c shows that the differences in the edge energies decreases as the system becomes W-rich. Whereas S-rich conditions thermodynamically favor zigzag edges (type A), intermediate or W-rich local conditions may promote the formation of the slanted edges (type B). In order to produce a coalesced monolayer with minimal bilayer coverage, the $WS_2$ analyzed here is synthesized in a multistep growth process. To achieve this, the temperature and metal precursor flux are controllably modulated over the course of the reaction.[29] This means that the growing film experiences multiple kinetic regimes and variable metal precursor concentrations. Whereas simulations indicate that type B edges grow quickly and would disappear in favor of zigzag edges if the reaction proceeded long enough, the growth conditions and the resulting experimental observations clearly indicate that the film has a kinetically driven morphology. Also, the $WS_2$ is grown in an $H_2$ ambient, which may promote a competition between growth and etching when the W precursor flow rate is modulated. This may lead to trapping of the faster-growing, experimentally observed slanted edges (~15°), even if not the fastest-growing edges predicted (19°). Therefore, adjusting the growth rate (by adjusting the gas flow) and/or controlling the nucleation density on the substrate could promote disappearance of fast-growing, slanted edges and provide routes to reduce the disregistry between growing $WS_2$ islands.

In summary, we have identified the nearly single-orientation character of coalesced monolayer WS$_2$ films grown by MOCVD and have uncovered the origin of unexpected, apparently irregular facets in their defect arrays. By connecting the film microstructure across length scales (e.g., atomic structure to multi-micrometer), we have created links that provide unexpected insights into the mechanism for crystal growth. Strikingly, we quantitatively characterize these irregular GBs to have very specific, well-defined orientations off the zigzag edges due to their kinetically driven growth conditions. This investigation utilized the combination of S/TEM imaging and ReaxFF MD simulations to uncover and quantify the relative orientations of GBs, demonstrating that these defect structures exist as subtle but widespread imperfections, despite growth conditions that produce nearly single-orientation, monolayer films. The defect arrays observed originate from the sub-unit-cell translational (longitudinal and transverse) offsets between the two neighboring grains with no angular misorientation as they coalesce. Further engineering of the growth process and selection of an ideal substrate can minimize this offset and lead to the epitaxial growth of wafer-scale, monolayer, single-crystalline 2D TMD films with little to no presence of such defect arrays.

**Acknowledgments**

This work was financially supported by the National Science Foundation (NSF) through the Pennsylvania State University 2D Crystal Consortium–Materials Innovation Platform (2DCC-MIP) under NSF cooperative agreement DMR-1539916. D.R.H., S.B., and N.A. acknowledge support from the NSF CAREER program (DMR-1654107) and the NSF program EFRI 2-DARE (EFRI-1433378). This work utilized resources provided by the NSF-MRSEC-sponsored Materials Characterization Lab at Penn State.